\documentclass{egpubl}
\usepackage{eurovis2025}

\EuroVisShort  

\usepackage[T1]{fontenc}
\usepackage{dfadobe}

\usepackage{cite}  
\BibtexOrBiblatex
\electronicVersion
\PrintedOrElectronic
\ifpdf \usepackage[pdftex]{graphicx} \pdfcompresslevel=9
\else \usepackage[dvips]{graphicx} \fi

\usepackage{egweblnk}

\usepackage{xspace}
\usepackage{booktabs}
\usepackage{multirow}

\newcommand{\textsand}[1]{{#1}}

\usepackage{longfbox}
\usepackage{colortbl}

\definecolor{appleredlight}{RGB}{255, 105, 97}
\definecolor{appleorangelight}{RGB}{255, 179, 64}
\definecolor{appleyellowlight}{RGB}{255, 212, 38}
\definecolor{applegreenlight}{RGB}{48, 219, 91}
\definecolor{applemintlight}{RGB}{102, 212, 207}
\definecolor{appleteallight}{RGB}{93, 230, 255}
\definecolor{applecyanlight}{RGB}{112, 215, 255}
\definecolor{applebluelight}{RGB}{64, 156, 255}
\definecolor{appleindigolight}{RGB}{125, 122, 255}
\definecolor{applepurplelight}{RGB}{218, 143, 255}
\definecolor{applepinklight}{RGB}{255, 100, 130}
\definecolor{applebrownlight}{RGB}{181, 148, 105}

\definecolor{applerednormal}{RGB}{255, 69, 58}
\definecolor{appleorangenormal}{RGB}{255, 159, 10}
\definecolor{appleyellownormal}{RGB}{255, 214, 10}
\definecolor{applegreennormal}{RGB}{48, 209, 88}
\definecolor{applemintnormal}{RGB}{99, 230, 226}
\definecolor{appletealnormal}{RGB}{64, 200, 224}
\definecolor{applecyannormal}{RGB}{100, 210, 255}
\definecolor{applebluenormal}{RGB}{10, 132, 255}
\definecolor{appleindigonormal}{RGB}{94, 92, 230}
\definecolor{applepurplenormal}{RGB}{191, 90, 242}
\definecolor{applepinknormal}{RGB}{255, 55, 95}
\definecolor{applebrownnormal}{RGB}{172, 142, 104}

\definecolor{applegrey}{RGB}{99, 99, 102}

\newfboxstyle{patternparamrad}{padding-top=2pt, padding-bottom=2pt, padding-left=4pt, padding-right=4pt, border-style=solid, height=6.5pt, border-radius=5pt}

\newfboxstyle{patternparam}{padding-top=2pt, padding-bottom=2pt, padding-left=3pt, padding-right=3pt, border-style=none, height=6.5pt}

\def\Pioneer{\lfbox[patternparam, background-color=applepinknormal!20]{{\normalfont \textsand{Pioneer}}}\xspace}

\def\Pioneers{\lfbox[patternparam, background-color=applepinknormal!20]{{\normalfont \textsand{Pioneers}}}\xspace}

\def\Pioneerbf{\lfbox[patternparam, background-color=applepinknormal!20]{{\normalfont \textsand{\textbf{Pioneer}}}}\xspace}

\def\Judge{\lfbox[patternparam, background-color=appleindigonormal!20]{{\normalfont \textsand{Judge}}}\xspace}

\def\Judgebf{\lfbox[patternparam, background-color=appleindigonormal!20]{{\normalfont \textsand{\textbf{Judge}}}}\xspace}

\def\Instructor{\lfbox[patternparam, background-color=applegreennormal!20]{{\normalfont \textsand{Instructor}}}\xspace}

\def\Instructorbf{\lfbox[patternparam, background-color=applegreennormal!20]{{\normalfont \textsand{\textbf{Instructor}}}}\xspace}

\def\Explorer{\lfbox[patternparam, background-color=applegrey!20]{{\normalfont \textsand{Explorer}}}\xspace}

\def\Explorerbf{\lfbox[patternparam, background-color=applegrey!20]{{\normalfont \textsand{\textbf{Explorer}}}}\xspace}

\def\Explainer{\lfbox[patternparam, background-color=appleyellownormal!20]{{\normalfont \textsand{Explainer}}}\xspace}

\def\Explainerbf{\lfbox[patternparam, background-color=appleyellownormal!20]{{\normalfont \textsand{\textbf{Explainer}}}}\xspace}

\def\Architect{\lfbox[patternparam, background-color=applebrownnormal!20]{{\normalfont \textsand{Architect}}}\xspace}

\def\Architectbf{\lfbox[patternparam, background-color=applebrownnormal!20]{{\normalfont \textsand{\textbf{Architect}}}}\xspace}

\renewcommand{\paragraph}[1]{
\noindent
\textbf{#1.}
}

\newcommand{\revise}[1]{{#1}}

\title[Exploring High-Dimensional Backstage]%
      {\revise{Navigating} High-Dimensional Backstage: A Guide for Exploring Literature for the Reliable Use of Dimensionality Reduction}

\author[Jeon et al.]
{\parbox{\textwidth}{
\centering 
    Hyeon Jeon$^{1}$\orcid{0000-0002-9659-2922}\thanks{hj@hcil.snu.ac.kr}, 
    Hyunwook Lee$^{2}$\orcid{0000-0002-5506-7347}, 
    Yun-Hsin Kuo$^{3}$\orcid{0009-0000-1891-8993},
    Taehyun Yang$^{1}$\orcid{0009-0000-1023-5266}, \\ 
    Daniel Archambault$^{4}$\orcid{0000-0003-4978-8479},
    Sungahn Ko$^5$\orcid{0000-0002-7410-5652},
    Takanori Fujiwara$^6$\orcid{0000-0002-6382-2752},
    Kwan-Liu Ma$^3$\orcid{0000-0001-8086-0366},
    Jinwook Seo\thanks{jseo@snu.ac.kr, Corresponding Author}$^{1}$\orcid{0000-0002-7734-822X
}
        }
        \\
{\parbox{\textwidth}{\centering $^1$Seoul National University, Republic of Korea $\quad$
         $^2$UNIST, Republic of Korea $\quad$ 
         $^3$University of California, Davis, CA $\quad$ \\ 
         $^4$Newcastle University, UK $\quad$
         $^5$POSTECH, Republic of Korea $\quad$
         $^6$Link\"oping University, Sweden \\ 
       }
}
}

\begin{document}


\maketitle
\begin{abstract}
   Visual analytics using dimensionality reduction (DR) can easily be unreliable for various reasons, e.g., inherent distortions in representing the original data. 
The literature has thus proposed a wide range of methodologies to make DR-based visual analytics reliable. 
However, the diversity and extensiveness of the literature can leave novice analysts and researchers uncertain about where to begin and proceed. 
To address this problem, we propose a guide for reading papers for reliable visual analytics with DR. 
Relying on the previous classification of the relevant literature, our guide helps both practitioners to (1) assess their current DR expertise and (2) identify papers that will further enhance their understanding. 
Interview studies with three experts in DR and data visualizations validate the significance, comprehensiveness, and usefulness of our guide.


\end{abstract}  

\section{Introduction}

Dimensionality reduction (DR) is widely used and also extensively researched. 
Over the past decades, researchers have introduced diverse DR techniques \cite{moor20icml, jeon22vis, joia11tvcg}, quality metrics \cite{jeon21tvcg, jeon24tvcg, zhang23tvcg}, and visual analytics systems that leverage DR \cite{choo13vda, chatzimparmpas20tvcg, sacha17tvcg}. Furthermore, the visualization community has produced several survey papers that look at DR literature from various perspectives, e.g., interaction with DR \cite{sacha17tvcg} or the evaluation of DR techniques \cite{espadoto21tvcg}.

Despite its usefulness, visual analytics using DR can easily be unreliable \cite{jeon25chi, nonato19tvcg}.
For example, DR techniques themselves intrinsically introduce distortions in representing the original data \cite{lespinats11cgf}. Also, DR projections often lack interpretability as they do not demonstrate the original high-dimensional features \cite{faust19tvcg}.
Such unreliability can lead analysts to draw incorrect insights from the data, which may cascade into downstream analyses and negatively affect overall decision-making \cite{jeon25chi}. The visualization and machine learning community has thus proposed diverse methodologies to alleviate this reliability problem \cite{jeon21tvcg, faust19tvcg, joia11tvcg, lespinats11cgf}.

However, though such methodologies exist, the breadth of the literature makes it difficult to understand and apply these methodologies in practice. 
Existing surveys \cite{ nonato19tvcg, jeon25chi} provide a broad overview of the literature related to the reliable use of DR. 
Nonetheless, they do not provide actionable steps for analysts and researchers when confronted with a DR problem, highlighting the need for guidance tailored to various levels of expertise.

To address this gap, we introduce the guide for interfacing with the literature on the reliable use of DR. 
\revise{Relying on the previous survey on DR literature about the reliable use of DR \cite{jeon25chi},} our guide provides a checklist that helps analysts to evaluate their expertise level and recommends a list of papers that align with their level, informed by the existing classification of the relevant literature \cite{jeon25chi}. 
Our guide is evaluated through expert interviews with three experts in the field.
We hope our guide will act as a valuable compass for practitioners at varying levels, facilitating more engagement and discussions in the relevant field.

\section{Background: Classification of the Literature}

\label{sec:background}

\revise{We want our guide to help analysts resolve 'practical problems' that occur within the visual analytics process using DR. 
We thus decide to rely on a classification of relevant literature derived from a previous survey \cite{jeon25chi}.
This classification organizes papers according to the visual analytic stage targeted, the specific challenges addressed, and the solutions to address the challenges. 
We therefore view these categories as an effective roadmap for exploring the literature on the reliable use of DR in visual analytics. The followings are the details of these classes.
}



\paragraph{\Pioneerbf}
The papers in this class ``pioneer'' new DR techniques to enhance the reliability of DR techniques themselves. This is done by (1) improving existing DR techniques \cite{jeon22vis, pezzotti16cgf} or (2) designing DR techniques from scratch \cite{joia11tvcg, fadel15neurocomp}. By doing so, these papers make DR projections to better represent the structure of original high-dimensional data. The class can be again divided into two types: \textsand{static} and \textsand{interactive}. While \textsand{static} \Pioneers contribute DR techniques that are not updated after the initial projection \cite{jeon22vis, desilva02neurips}, \textsand{interactive} \Pioneers propose DR techniques that interactively update the inner logic based on user input \cite{joia11tvcg}.

\paragraph{\Judgebf}
These papers aim to enhance the evaluation of DR projections. This is done by proposing new evaluation metrics \cite{jeon24tvcg, jeon21tvcg, sips09cgf}, workflows \cite{aupetit14beliv}, and libraries \cite{jeon23vis}. The papers thus contribute to making visual analytics using DR projections more accurate. 

\paragraph{\Instructorbf}
The papers in this class provide a high-level overview of DR literature. They contribute literature reviews \cite{nonato19tvcg, vandermaaten09jmlr} or benchmark experiments \cite{xia22tvcg, espadoto21tvcg, etemadpour15tvcg} in which practitioners can use as reference in selecting appropriate DR techniques or quality metrics that align with their analytic tasks. These papers share a similar goal with our paper---to educate practitioners about using DR more reliably. 
However, our contribution differs in terms of supporting practitioners to self-examine their expertise and suggesting tailored recommendations.

\paragraph{\Explorerbf}
These papers contribute visual analytics systems or novel visualizations that enhance the exploration of high-dimensional subspaces. The papers mainly tackle a scalability problem that the number of subspaces to investigate increases as dimensionality increases \cite{nam13tvcg, yi05ivis}. 

\paragraph{\Explainerbf}
The papers in this class explain high-dimensional features not well represented in 2D projection, enhancing the interpretability of DR-based visual analytics. This class can be again divided into two types: \textsand{distortion} and \textsand{attribute}. While \textsand{distortion} \Explainer aims to show where and how much projections are distorted in representing the original data \cite{lespinats11cgf}, \textsand{attribute} \Explainer shows high-dimensional attribute values along with the projections \cite{faust19tvcg}. 

\paragraph{\Architectbf}
The papers in this class mostly contribute visual analytics systems that enhance users' understanding of specific DR techniques (e.g., t-SNE \cite{chatzimparmpas20tvcg}) or help users to overcome DR techniques' inaccuracy of instability, thereby conducting more reliable visual analytics \cite{xia17tvcg}. 

\section{Methodology} 


\paragraph{Authors}
Nine visualization researchers, who are the authors of this paper, participate in curating the guide. 
The authors consist of two Ph.D. students, one undergraduate student, two postdoctoral researchers, and four professors. 
The authors' expertise also spans visualization, human-computer interaction, and machine learning.
These nine authors are also the authors of the survey paper that proposes the classification of the related literature (\autoref{sec:background}) \cite{jeon25chi}.

\paragraph{Procedure}
hree authors (one Ph.D. student, one postdoctoral researcher, and one professor) engage in formulating the initial version of our guide. These authors discuss its structure by conducting two online Zoom meetings. 
Then, the student, who is the main author of this paper, create the initial draft. 
The three authors met again to confirm the draft. 
Then, the authors conduct online meetings to review and revise our guide. The main author participate in every meeting to explain the current version of our guide and to lead the meeting. Other authors participated irregularly.
The meetings are iteratively conducted until all authors agree on the structure. In total, ten meetings are conducted from August 2024 to January 2025. 

\paragraph{\revise{Design Rationale}}
\revise{Our design rationales aims to make our guide more approachable to broader audiences. 
First, we want our guide to \textbf{(R1) support analysts with diverging level of expertise.}
This is because DR is widely used by both visualization experts and also analysts in other domains (e.g., bioinformatics or business \cite{cashman25arxiv}) who have low literacy in visualizations and DR.
For the same reason, we want our guide to \textbf{(R2) be less burdensome to be leveraged by analysts}.
}



\section{Our Guide for Exploring DR Literature}

\label{sec:guide}

\begin{figure}
    \centering
    \includegraphics[width=\linewidth]{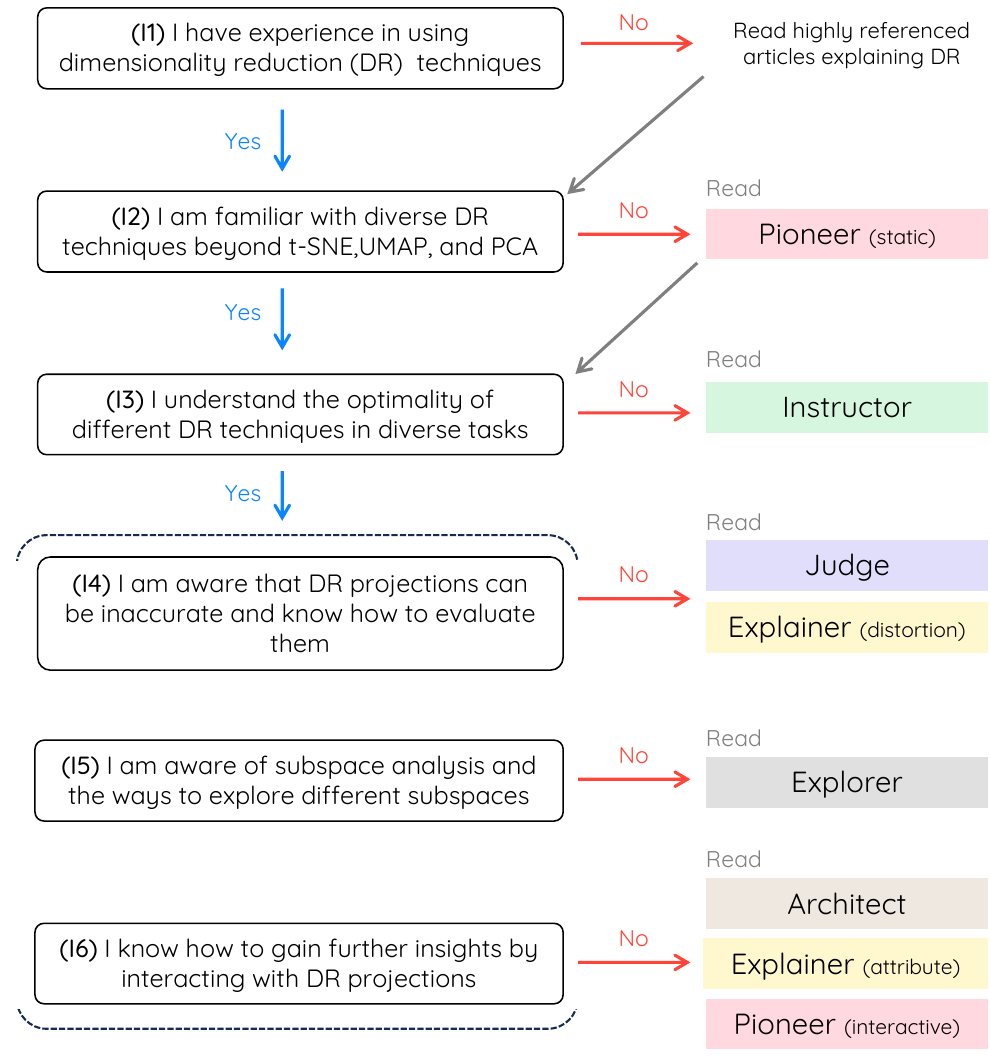}
    \vspace{-5mm}
    \caption{The flow diagram illustrating our guide. This guide (\autoref{sec:guide}) is structured to follow the level of experience with DR. 
 }
    \label{fig:checklist}
\end{figure}

We introduce our guide to explore the literature on making DR-based visual analytics more reliable (\autoref{fig:checklist}). 
\revise{Although the previous survey on DR that we rely on \cite{jeon25chi} provides a comprehensive ``map'' of the literature, practitioners may still struggle to identify which parts directly address their specific problems.
To address this problem,} our guide includes a checklist with six items that help analysts assess their expertise level, with each item corresponding to a recommendation of a cluster of papers to read to enhance their expertise (R1). Our guide and checklists can thus be interpreted as our suggested order for reading papers to \revise{navigate and understand} the relevant field comprehensively.

The first part of our guide is structured by increasing levels of expertise with the literature (\textbf{I1}, \textbf{I2}, and \textbf{I3}): if an analyst can affirmatively answer a preceding checklist item, they should proceed to the next. If not, we suggest the analyst reference the relevant papers from our list. Once sufficient expertise in DR is attained, analysts should focus on the specific checklist items that best align with their analytic contexts and tasks (\textbf{I4}, \textbf{I5}, and \textbf{I6}). 
\revise{The first three items (I1--I3) are the preliminaries to readily understand the papers on these advanced topics. }
\revise{We use simple yes-no questions to reduce the efforts of analysts in following the guide (R2).}
Below is a detailed explanation of each checklist item (\textbf{I1}--\textbf{I6}). 
Please refer to our interactive web guide\footnote{\url{https://dr-reliability.github.io/guide/}} for the detailed list of papers corresponding to each checklist item.

\paragraph{(I1) I have experience in using DR techniques}  Analysts who do not meet this checklist item should first refer to basic articles explaining widely used DR techniques and their functionalities before accessing the papers in our list. 
We recommend to ``overview'' the DR literature by reading highly referenced articles explaining widely used DR techniques. This includes the surveys on DR techniques \cite{espadoto21tvcg, cunningham15jmlr} and technical papers that explain famous DR techniques (e.g., $t$-SNE \cite{maaten08jmlr, maaten14jmlr}, Autoencoder \cite{hinton06science}).
We also recommend reading the user guides of DR libraries\footnote{\url{https://umap-learn.readthedocs.io/en/latest/basic_usage.html}}\footnote{\url{https://scikit-learn.org/stable/modules/unsupervised_reduction.html}}, as they are closely tied to practical analysis.
An analyst can proceed to \textbf{I2} if the one satisfies this checklist item.

\paragraph{(I2) I am familiar with diverse DR techniques beyond $t$-SNE, UMAP, and PCA}  Without meeting this checklist item, analysts will likely lack the knowledge required to choose a DR technique that suits their tasks. The analysts should first be aware that various DR techniques exist, each with different purposes and focus. We thus recommend referring to the \textsand{static} \Pioneer group that proposes DR techniques producing static projections. Reviewing these papers will inform the analysts that famous techniques are not always optimal and that alternative techniques exist to complement them. Analysts can proceed to \textbf{I3} if they meet this checklist item.

\paragraph{(I3) I understand the optimality of different DR techniques for diverse tasks}  Although analysts may be aware that different tasks have optimal techniques, mapping tasks to the appropriate techniques cannot be easily accomplished by reading individual papers. This is because each paper often asserts that its techniques are the best in various aspects, which is sometimes misleading. We thus strongly suggest analysts refer to papers in the \Instructor group, where they offer objective evaluations or comparisons of different techniques. After saying `yes' to this checklist item, analysts can examine their expertise level for advanced topics (\textbf{I4}, \textbf{I5}, and \textbf{I6}).

\paragraph{(I4) I know DR projections can be inaccurate and know how to evaluate them}  Even with the proper selection of DR techniques, resulting projections can be less optimal or inaccurate in representing the structure of the original data. Therefore, if analysts want to improve the accuracy of their projections, we recommend focusing more closely on evaluation metrics, aiming to identify reliable projections. This can be done by referencing the papers in \Judge group as they propose various evaluation metrics. We also suggest reading \textsand{distortion}-type \Explainer papers that address inaccuracy by being more aware of inherent projection distortions.

\paragraph{(I5) I am aware of subspace analysis and the ways to explore different subspaces}  If analysts do not meet the checklist item, referring to \Explorer papers is recommended. These papers contribute various visualization methodologies that scalably support subspace analysis, enabling analysts who read these papers to meet the checklist item.

\paragraph{(I6) I know how to gain further insights by interacting with DR projections}  This checklist item asks analysts whether they have enough operational knowledge to make their analysis interpretable and to reflect their domain knowledge. In terms of interpretability, we suggest analysts reference \Architect type papers and \Explainer type papers that deal with interpretability (\textsand{attribute}-type). To leverage domain knowledge, we recommend consulting \textsand{interactive} \Pioneer papers. These papers offer interactive DR techniques that dynamically modify their algorithm in response to user input, thus integrating domain expertise.

\begin{table*}[t]
    \centering
    
    \caption{\textit{The questionnaires for our expert interview.} The rating of our guide based on these questions is available in \autoref{tab:scores}. Overall, the experts agree on the significance, comprehensiveness, and usefulness of our guide.}
    \scalebox{1.00}{
    \begin{tabular}{l|ll}
    \toprule
        \textbf{Criteria} & \textbf{No. }& \textbf{Questions}\\
    \midrule
        \multirow{2}{*}{Significance} &  Q1 & \textit{The guide addresses critical problems in DR-based visual analytics research and practice.}\\
        & Q2 & \textit{Adopting this guide can substantially influence and improve the current practices in the relevant field. }\\
    \midrule
        \multirow{2}{*}{Comprehensiveness} & Q3 & \textit{The guide covers the essential range of topics and challenges in the DR-based visual analytics.} \\
        & Q4 & \textit{The guide provides sufficient detail and depth to guide both novice and experienced analysts.} \\ 
    \midrule
    \multirow{2}{*}{Usefulness} & Q5 & \textit{The guide offers clear, actionable steps that practitioners can readily adopt to increase their expertise.} \\
    & Q6 & \textit{Following the guide will help analysts properly use DR in different analytical contexts.} \\ 
        
    \bottomrule
    \end{tabular}
    }
    \label{tab:questions}
\end{table*}

\section{Expert Evaluation}

\label{sec:expert}

We evaluate our guide with expert interview.

\subsection{Study Design}

\paragraph{Participants}
We want the experts to have sufficient expertise and experience in DR and also in data visualization. We thus recruit experts who satisfy three conditions: (1) regularly publish visualization papers in visualization, human-computer interaction, or machine learning venues; (2) previously have published visualization papers related to DR; and (3) hold Ph.D. degree. 
Our experts consist of two professors and one postdoctoral researcher (three in total), who have 12.6 years of experience in visualization research on average. \revise{Note that the list of the experts do not overlap with the authors of our guide.}

\paragraph{Interview procedure}
We interview each expert individually.
One experimenter manages all interviews. 
After the experts consent to participate in the experiment, the interviewer explains the guide and checklist in detail for 10 minutes. 
Then, the experts can freely ask questions about the guide.
Finally, the interviewer asks experts to assess the significance, comprehensiveness, and usefulness of our guide on a Likert scale and provide the reasoning (see \autoref{tab:questions}).
All interviews finished within 40 minutes.

\subsection{Results}

We detail the results of our expert interview. The experts agree on the significance, comprehensiveness, and usefulness of our guide. The interview also discovers interesting future directions.

\newcommand{\accolor}[1]{\cellcolor{appleredlight!#1}}

\begin{table}[t]
    \centering
    \setlength{\tabcolsep}{6.8pt}
    \caption{The expert ratings on significance (Sig.), comprehensiveness (Comp.), and usefulness (Use.) of our guide (1: strongly disagree, 5: strongly agree). We apply a red highlight to indicate agreement, with an opacity scale ranging from 0\% (neutral, score 3) to 100\% (strong agreement, score 5), and a linear gradient filling the values in between.}
    \begin{tabular}{r|cccccc|c}
\toprule
& \multicolumn{2}{c}{\textbf{Sig. }} & \multicolumn{2}{c}{\textbf{Comp.}} & \multicolumn{2}{c|}{\textbf{Use.}} & \multirow{2}{*}{\textbf{Avg.}} \\ 
\cmidrule(lr){2-3} \cmidrule(lr){4-5} \cmidrule(lr){6-7} 
 &   \textbf{Q1} & \textbf{Q2} & \textbf{Q3} & \textbf{Q4} & \textbf{Q5} & \textbf{Q6} \\
 \midrule
   \textbf{P1}  &  5 \accolor{100}& 5 \accolor{100}& 3 & 4 \accolor{50}& 4 \accolor{50}& 4 \accolor{50} & 4.17 \accolor{58}\\
   \textbf{P2} & 4 \accolor{50}& 5 \accolor{100}& 4 \accolor{50}& 4 \accolor{50}& 5 \accolor{100}& 3 & 4.17 \accolor{58}\\
   \textbf{P3} & 4 \accolor{50}& 5 \accolor{100}& 5 \accolor{100}& 4\accolor{50}& 5\accolor{100}& 4 \accolor{50}& 4.50 \accolor{75}\\
   \midrule
   \textbf{Avg.} & 4.33 \accolor{67}& 5.00 \accolor{100}& 4.00 \accolor{50}& 4.00\accolor{50}& 4.66 \accolor{83}& 3.66 \accolor{33} & 4.28 \accolor{64}\\
   \bottomrule
    \end{tabular}
    \label{tab:scores}
\end{table}

\paragraph{Significance} 
The experts overall agree on the significance of the guide, providing scores higher than 4 (agree) on average for both questions. The experts acknowledge that the guideline addresses important challenges in DR-based visual analytics. They emphasize its potential to positively influence current practices, particularly for beginners. 

\paragraph{Comprehensiveness}
As with the previous criteria, the experts agree with the comprehensiveness of the guide. 
However, qualitative results indicate that more efforts are needed to make the guide more comprehensive. Aligned to this quantitative result, the experts point out potential gaps for complete novices in DR. 
Investigating the difficulty of the papers we recommend and building a more fine-grained guide will be an interesting future direction to explore.

\paragraph{Usefulness}
The experts provide scores higher than 4 (agree) on average, indicating their agreement with the usefulness of our guide.
The experts found the guideline highly actionable, praising its checklist items and web interface. They remark on how it provides a structured approach to learning and using DR effectively. 

\section{\revise{Discussions}}

\subsection{\revise{Extending Our Approach to Other Surveys}}

\revise{
We rely on an existing classification of DR literature to establish our guide.
The rationale here is that survey papers already provide comprehensive references for practitioners to address their challenges and for researchers to draw upon in their studies (\autoref{sec:background}). Our evaluation (\autoref{sec:expert}) verifies the appropriateness of this approach in building an effective guide for exploring the literature.
}

\revise{
It is therefore natural to extend our approaches to other survey papers in the visualization domain.
One possible direction is to combine multiple surveys into a single guide.
For example, the DR literature includes several survey papers offering different perspectives \cite{espadoto21tvcg, nonato19tvcg, jeon25chi}.
Integrating these surveys into a unified guide would reduce the burden on practitioners and expose them to diverse viewpoints.
Another promising direction is to automate the construction of the guide from survey papers, making it easier for practitioners to access insights from diverse domains.
}

\subsection{Limitations}

Although verified to be effective (\autoref{sec:expert}), our guide has several limitations. 
\revise{First, checklist items that self-reports expertise of practitioners may be less reliable than probing their actual knowledge.
For example, practitioners may underestimate their knowledge, leading to unnecessary reading of papers.
An alternative is to incorporate two or three questions assessing DR knowledge, enabling practitioners to evaluate their expertise more objectively.
Another limitation is that our paper recommendations remain coarse-grained, with each category comprising dozens of papers.
Providing more fine-grained checklist items and recommendations may enhance the guideline's usability.
}
Finally, our guide is built upon the literature available to date. As the literature on DR-based visual analytics continues to evolve, the long-term generalizability of our framework remains uncertain. 
Investigating how the relevant field has altered over time will be an interesting future research avenue.


\section{Conclusion}

While DR-based visual analytics may be prone to unreliability, there is a lack of comprehensive guidance on how novice analysts and researchers can explore the literature to improve their analyses.
We contribute a guide for reading papers in the relevant literature to address this problem. 
By providing both a self-assessment checklist and curated paper recommendations, our guide functions as an actionable resource to strengthen expertise in the field.
Three expert researchers in the relevant field confirm the significance, comprehensiveness, and usefulness of our guide. 

\bibliographystyle{eg-alpha-doi} 
\bibliography{ref}       


\newpage


\end{document}